\documentclass[12pt]{article}

\usepackage{amssymb,amsmath}
\usepackage{color}
\usepackage{graphicx}
\usepackage{tikz}
\usepackage{algorithm}
\usepackage{algpseudocode}
\usepackage{epstopdf}
\usepackage{url}
\usepackage{rotating}

\def\Abb{{\mathbb A}}
\def\Bbb{{\mathbb B}}
\def\Cbb{{\mathbb C}}

\def\R{{\mathbb R}}

\def\l{\lambda}
\def\L{\Lambda}

\def\s{\sigma}

\def\th{\theta}

\def\bmu{{\boldsymbol \mu}}
\def\bnu{{\boldsymbol \nu}}
\def\bth{{\boldsymbol \theta}}

\def\ap{\rightarrow}

\def\seq{\subseteq}

\def\fa{\; \forall}

\def\nm{\Vert}

\renewcommand{\and}{\mbox{$\wedge$}}

\newcommand{\bc}{\begin{center}}
\newcommand{\ec}{\end{center}}
\newcommand{\be}{\begin{equation}}
\newcommand{\ee}{\end{equation}}
\newcommand{\bd}{\begin{displaymath}}
\newcommand{\ed}{\end{displaymath}}
\newcommand{\ba}{\begin{array}}
\newcommand{\ea}{\end{array}}
\newcommand{\ben}{\begin{enumerate}}
\newcommand{\een}{\end{enumerate}}
\newcommand{\bit}{\begin{itemize}}
\newcommand{\eit}{\end{itemize}}
\newcommand{\beq}{\begin{eqnarray}}
\newcommand{\eeq}{\end{eqnarray}}
\newcommand{\btab}{\begin{tabular}}
\newcommand{\etab}{\end{tabular}}
\newcommand{\bfig}{\begin{figure}}
\newcommand{\efig}{\end{figure}}

\newtheorem{defi}{Definition}
\newtheorem{thm}{Theorem}
\newtheorem{lem}{Lemma}
\newcommand{\bdefi}{\begin{defi}}
\newcommand{\edefi}{\end{defi}}
\newcommand{\bthm}{\begin{thm}}
\newcommand{\ethm}{\end{thm}}
\newcommand{\blem}{\begin{lem}}
\newcommand{\elem}{\end{lem}}

{\bf}{\it}
{\bf}{\it}
{\bf}{\rm}
{\bf}{\it}
\newtheorem{theorem}{Theorem}[section]
{\bf}{\it}
{\bf}{\it}
{\bf}{\rm}

\definecolor{verm}{rgb}{0.6,0.2,0.2}
\definecolor{purp}{rgb}{0.3,0.1,0.6}
\definecolor{purple}{rgb}{0.4,0.0,0.6}
\definecolor{bggreen}{rgb}{0.1,0.3,0.1}
\definecolor{dgreen}{rgb}{0.1,0.6,0.1}
\definecolor{black}{rgb}{0.0,0.0,0.0}
\definecolor{crim}{rgb}{0.3,0.1,0.1}
\definecolor{dred}{rgb}{0.5,0.1,0.1}

\begin{document}

\title{
Inferring Genome-Wide Interaction Networks Using the
Phi-Mixing Coefficient, and Applications to Lung Cancer
}
\author{
Nitin Singh, Mehmet Eren Ahsen, Shiva Mankala,\\
Hyun-Seok Kim,
Michael A.\ White and M.\ Vidyasagar
\thanks{
NS is with Apple R\&D, Austin, TX; MEA is with IBM Thomas J.\ Watson
Research Center, Yorktown Heights, NY;
SM is with High Mark Inc., Pittsburgh, PA; Hyun-Seok Kim is with
Yonsei University College of Medicine, Seoul, Korea;
MAW is with the UT Southwestern Medical Center, Dallas, TX;
and MV is with the University of Texas at Dallas, Richardson, TX.
Corresponding author: M.\ Vidyasagar; email:m.vidyasagar@utdallas.edu.
}
}

\maketitle

\begin{abstract}

Constructing gene interaction networks (GINs) from high-throughput 
gene expression data is an important and challenging problem in systems biology.
Existing algorithms produce networks that either have undirected
and unweighted edges, or else are constrained to contain no cycles,
both of which are biologically unrealistic.
In the present paper we propose a new algorithm, based on a concept from
probability theory known as the phi-mixing coefficient, that produces
networks whose edges are weighted and directed, and are permitted to
contain cycles.
Because there is no ``ground truth'' for genome-wide networks on a human
scale, we analyzed the outcomes of
several experiments on lung cancer, and matched
the predictions from the inferred networks with experimental results.
Specifically, we inferred three networks 
(NSCLC, Neuro-endocrine NSCLC plus SCLC, and normal)
from the gene expression measurements of 157 lung cancer and 59
normal cell lines;
compared with the outcomes of
siRNA screening of 19,000+ genes on 11 NSCLC cell lines;
and analyzed data from
a ChIP-Seq experiment to determine putative downstream
targets of the lineage specific oncogenic transcription factor ASCL1.
The inferred networks displayed a scale-free
or power law behavior between the degree
of a node and the number of nodes with that degree.
There was a strong correlation between the degree of a gene in the
inferred NSCLC network and its essentiality for the survival of the cells.
The inferred downstream neighborhood genes of ASCL1
in the SCLC network were significantly
enriched by ChIP-Seq determined putative target genes, while
no such enrichment was found in the inferred NSCLC network.

\end{abstract}

%
%

\section{Introduction}\label{sec:intro}

The behavior of cells is governed by complex interactions amongst
genes and gene products.
Diseases such as cancer have their origin in a departure of these
interactions from their normal patterns.
By comparing the interactions that are present in normal cells versus cancerous
cells, or between cells manifesting
different forms of cancer, it is possible to
draw some conclusions about the triggers of cancer and/or potential
therapeutic targets.

Interactions and regulatory mechanisms that
prevail between genes can be studied a few genes at a time,
under carefully controlled experimental conditions.
These interactions can then be aggregated to construct
larger networks.
However, such an approach is subject to two potential pitfalls.
First, even aggregations of multiple public and commercial
databases rarely contain more than 10,000 genes, whereas the human genome
contains around 22,000 genes.
The main reason for this is that biologists tend to study genes
that are ``interesting,'' with the result that ``uninteresting'' genes
and their neighborhoods do not get explored.
The second potential pitfall is that the interactions in the aggregated
network may be present under widely different, and perhaps even 
contradictory, experimental conditions;
thus aggregating all these interactions into a common network may not
be justifiable.

To overcome these potential pitfalls, in recent years there has been
interest in reverse-engineering whole-genome interaction networks
from simultaneous measurements of the expression levels of \textit{all}
(or at least most) genes in many samples, under
a common set of experimental conditions.
Such a network can be referred to as ``whole-genome context-dependent.''
In such an approach, the expression level of each gene is viewed as
a random variable, say $X_1$ through $X_n$ where $n \sim 20,000$
is the total number of genes whose measurements are taken.
Then one attempts to construct a network (or graph) consisting of
$n$ nodes, one for each gene,
such that an interaction between two nodes is represented
as an edge in the network between the corresponding nodes.
%

In order to make the theory work, assumptions are imposed on the underlying
network that do not always coincide with biological realism.
Existing methods for reverse-engineering whole-genome networks fall into
two broad categories, which might be termed ``Bayesian'' and
`information-based.''
In the Bayesian framework \cite{Friedman2000, BF02,Friedman04, Yu2004,KF09},
it is assumed that the network consists of directed edges and is acyclic.
In other words, if there exists a path from gene $A$ to gene $B$, there cannot
exist a path from gene $B$ back to gene $A$.
This contradicts the fact that ``real'' biological networks contain
a myriad of feedback loops in order to achieve stasis.
In information-based methods, of which ARACNE \cite{Basso2005}
and CLR \cite{Faith2007}
are among the best-known,
one computes a quantity called the mutual information between pairs of genes,
and then uses
a mathematical relationship known as the data processing inequality (DPI)
to determine which interactions are to be retained in the network \cite{Basso2005,Margolin2006,Jang2013}.
This approach can be proven to
recover the joint probability distribution of all random variables,
under the very restrictive assumption that the 
joint probability distribution
consists of only first order or second order terms.
The main drawback of this approach is that mutual information
is a \textit{symmetric} measure of the dependence between two random
variables.
Therefore networks constructed using information-based methods
are perforce \textit{undirected}.
When one of the genes is a known transcription factor, the lack
of directionality of the edges is not an isssue.
However, when the aim is to infer genome-wide interaction networks,
the lack of directionality is a serious shortcoming.
Some of the recently developed methods such as GENIE3 \cite{Huynh-Thu2010} and bLARS \cite{Singh2015a}
and ANOVA $\eta^2$ \cite{Kuffner2012} can infer directed weighted graphs.
However, to date these methods have been tested only on small organisms
such as E.\ coli or synthetic data sets, and
have not been assessed in the context of human genome-wide GIN.

Our motivation in devising yet another method for inferring interaction
networks was to address each of the issues raised above.
%
The algorithm presented here makes use of a \textit{directionally sensitive}
measure of dependence between random variables, known as the
phi-mixing coefficient; accordingly the algorithm is referred to as
``phixer.''
The networks produced by phixer contain edges that are both directed
as well as weighted (between $0$ and $1$), thus providing information
about both the direction and the strength of the interaction between
genes.

The phixer algorithm was used to reverse-engineer several
gene interaction networks (GINs) in the context of lung cancer.
These networks encompassed (i) non-small cell lung cancer (NSCLC),
(ii) small cell lung cancer (SCLC), (iii) both of the above, and (iv)
normal lung cells.
One of the major difficulties in validating an inferred genome-wide
GIN is that there is no ``ground truth'' network
against which it can be compared.
Therefore we used phixer-generated GINs to predict biological outcomes
from whole-genome scale data sets, and compared against the actual outcomes.
There was excellent agreement between the two, as described below.
Thus we believe that the phixer algorithm offers an attractive, state
of the art algorithm for inferring genome-wide GINs.


\section{Theoretical Details}

\subsection{Background}

In order to infer an interaction network from whole-genome expression data,
one of the most commonly used approaches is to view the
expression level of each gene 
as a random variable, and the measurements of the gene expression levels
as independent samples of that random variable.
Let $n$ denote the number of genes and $m$ denote the number of
samples.
Then the data is assumed to consist of $m$ statistically independent
samples of the joint random variable $(X_1 , \ldots , X_n)$.
There is no assumption that these random variables are independent;
indeed, the objective of the exercise is to determine their interdependence.

Suppose $\Abb$ is a finite set, say $\Abb = \{ 1 , \ldots ,
n \}$.\footnote{Actually we should write $\Abb = \{ a_1 , \ldots , a_n \}$,
because the elements of $\Abb$ are just labels and do not correspond to
integers.}
Suppose $X$ is a random variable assuming values in $\Abb$ with associated
probability distribution $\bmu$.
Thus $\mu_i = \Pr \{ X = i \}$.
Then the entropy of $X$, or the entropy of the probability distribution $\bmu$,
is defined by
\bd
H(X) = H(\bmu) = - \sum_{i=1}^n \mu_i \log \mu_i .
\ed
Now suppose $\Abb = \{ 1 , \ldots , n \} , \Bbb = \{ 1 , \ldots , m \}$
are finite sets, and that $X,Y$ are random variables assuming values in $\Abb$
and $\Bbb$ respectively.
Let $\bth$ denote the joint distribution of $(X,Y)$.
Thus $\bth$ is a probability distribution on the product set $\Abb \times 
\Bbb$.
Let $\bmu,\bnu$ denote the marginals of $\bth$ on $\Abb$ and $\Bbb$
respectively.
Thus $X$ has the probability distribution $\bmu$ and $Y$ has the
distribution $\bnu$.
With this notation, the mutual information between $X$ and $Y$ is defined as
\bd
I(X,Y) = H(X) + H(Y) - H(X,Y) = H(\bmu) + H(\bnu) - H(\bth) .
\ed
An alternate and equivalent expression for the mutual information is
\bd
I(X,Y) = \sum_{i=1}^n \sum_{j=1}^m \phi_{ij} \log \frac{ \phi_{ij} }
{ \mu_i \nu_j } .
\ed
Mutual information is always symmetric and nonnegative; that is,
$0 \leq I(X,Y) = I(Y,X) \leq \min \{ H(X) , H(Y) \}$.
Moreover, $I(X,Y) = 0$ if and only if $X$ and $Y$ are independent random
variables.

Suppose $X,Y,Z$ are random variables assuming values in finite sets
$\Abb,\Bbb,\Cbb$ respectively.
Then $X$ and $Z$ are said to be {\bf conditionally independent}
given $Y$, denoted by $(X \perp Z ) | Y$, if
for all $i \in \Abb , j \in \Bbb , k \in \Cbb$, it is true that
\bd
\Pr \{ X = i \& Z = k | Y = j \}
= \Pr \{ X = i | Y = j \} \cdot \Pr \{ Z = k | Y = j \} .
\ed
It is easy to show that the above relationship also implies that,
for all $S \seq \Abb , j \in \Bbb , U \seq \Cbb$, it is true that
\bd
\Pr \{ X \in S \& Z \in U | Y = j \} 
= \Pr \{ X \in S | Y = j \} \cdot \Pr \{ Z \in U | Y = j \} .
\ed
A very useful inequality, known as the ``data processing inequality''
or DPI for short,
states that whenever $(X \perp Z ) | Y$ the following inequality holds:
\be\label{eq:dpi-inf}
I(X,Z) \leq \min \{ I(X,Y), I(Y,Z) \} .
\ee
See \cite[p.\ 34]{Cover2006} for a detailed discussion.

One of the first papers to infer GINs using mutual information 
is \cite{Butte2000}.
In that paper, the authors compute the mutual information between every
pair of genes, and introduce an undirected edge between nodes $i$ and $j$ if
and only if the mutual information $I(X_i,X_j)$ between the corresponding
random variables $X_i$ and $X_j$ exceeds a minimum threshold.
They refer to the resulting (undirected) graph as an ``influence network.''
Indeed, in their framework, the presence of an (undirected)
edge between two nodes $i$ and $j$
makes no distinction between gene $i$ influencing gene $j$
or vice versa.
Also, no distinction is made between direct and indirect influence.
As a consequence, the influence networks produced by the method in
\cite{Butte2000} are extremely dense.

Algorithm for the Reconstruction of Accurate Cellular Networks (ARACNE)
\cite{Margolin2006} builds upon \cite{Butte2000} and produces 
networks that are not overly dense by pruning out large number of edges.
Specifically, for each triplet $i,j,k$, they compute all the three
mutual informations $I(X_i,X_j)$, $I(X_i,X_k)$ and $I(X_j,X_k)$.
Since the exact probability distributions are not known and only samples
are available, they use Gaussian kernel approximations for the
various joint distributions.
Then they identify the smallest amongst the three numbers and discard
the corresponding edge.
Thus if
\bd
I(X_i,X_k) \leq \min \{ I(X_i,X_j), I(X_j,X_k) \},
\ed
then they discard the edge between nodes $i$ and $k$.
The justification for the pruning strategy is that if
the joint distribution of all $n$ random variables has the form
\be\label{eq:22}
\phi(X_1 , \ldots , X_n) = \frac{1}{{\rm const.}}
\prod_{i=1}^n \psi(X_i) \cdot
\prod_{i,j = 1}^n \phi_{ij}(X_i,X_j) ,
\ee
then the algorithm produces the correct interaction graph.
However, the assumption that the joint probability distribution
has the above form is quite unrealistic.
Furthermore, the pruning strategy used in this algorithm implies
that the GIN generated will never contain a complete subgraph
of three nodes.
In other words, if there is an edge between nodes $i$ and $j$, and between
nodes $j$ and $k$, then there cannot be an edge between 
nodes $i$ and $k$.
But biology is full of small local networks that contain three-node
complete subgraphs.
Therefore new methods are required to generate
more biologically realistic interaction networks.

\subsection{Phi-mixing Coefficient as a Measure of Dependence}

The phixer algorithm
is based on computing the so-called $\phi$-mixing coefficient
between two random variables.
The $\phi$-mixing coefficient was introduced in \cite{Ibragimov62}
as a measure of the asymptotic long-term
independence of a stationary stochastic process,
and was used to prove laws of large numbers for non-i.i.d.\ processes.
The general definition \cite[Definition 2.1]{MV-03}
can be readily adapted to define a quantitative
measure of the dependence between two random variables \cite[page 3]{Doukhan94}.

If $X$ and $Y$ are random variables assuming values in 
possibly distinct finite\footnote{The assumption that both random variables
are finite-valued is made purely for convenience in exposition.
In the general case, the sets $S$ and $T$ would have to belong to the
$\s$-algebras generated by the random variables $X$ and $Y$ respectively,
and the maximum would have to be replaced by the supremum.}
sets $\Abb = \{ 1 , \ldots , n \}$ and $\Bbb = \{ 1 , \ldots , m \}$
respectively, the $\phi$-mixing coefficient $\phi(X|Y)$ is defined as
\be\label{eq:phi-def}
\phi(X|Y) := \max_{S \seq \Abb , Y \seq \Bbb} 
| \Pr \{ X \in S | Y \in T \} - \Pr \{ X \in S \} | .
\ee
Thus $\phi(X|Y)$ is the maximum difference between the conditional
and unconditional probabilities of an event involving only $X$,
conditioned over an event involving only $Y$.
It is well known that
the $\phi$-mixing coefficient has the following properties:
\ben
\item $\phi(X|Y) \in [0,1]$.
\item In general, $\phi(X|Y) \neq \phi(Y|X)$.
Thus the $\phi$-mixing coefficient gives directional information.
\item $X$ and $Y$ are independent random variables if and only if
$\phi(X|Y) = \phi(Y|X) = 0$.
\item The $\phi$-mixing coefficient is invariant under any one-to-one
transformation of the data.
Thus if $f: \Abb \ap \Cbb, g: \Bbb \ap {\mathbb D}$ are one-to-one
and onto maps, then
\bd
\phi(X|Y) = \phi(f(X) | g(Y)) .
\ed
\een

While (\ref{eq:phi-def}) is suitable for \textit{defining} the quantity
$\phi(X|Y)$, it cannot be directly used to \textit{compute it}.
Direct application of (\ref{eq:phi-def}) requires us to take the maximum
over all subsets of $\Abb$ and $\Bbb$, and would thus require
$2^{| \Abb | + | \Bbb |}$ computations.
A recent paper \cite{Ahsen2014} provides a closed form formula
for the computation of $\phi(X|Y)$, which implies that $\phi(X|Y)$ can
be calculated in polynomial time.
Let $\Theta \in [0,1]^{n \times m}$ denote the joint
distribution of $X$ and $Y$ written out as a matrix.
In other words,
\bd
\th_{ij} = \Pr \{ X = i \& Y = j \} , \fa i \in \Abb , j \in \Bbb .
\ed
Let $\bmu,\bnu$ denote the marginal distributions of $X$ and $Y$
respectively; thus
\bd
\mu_i = \Pr \{ X = i \} = \sum_{j=1}^m \th_{ij} ,
\nu_j = \Pr \{ Y = j \} = \sum_{i=1}^n \th_{ij} .
\ed
Define $\Psi \in [0,1]^{n \times m}$ as the outer product of $\bmu$ and
$\bnu$; thus
\bd
\psi_{ij} = \mu_i \nu_j , \fa i , j .
\ed
Then $\Psi$ is a rank one matrix, and is the joint distribution 
that $X$ and $Y$ would have if they were independent.
Define $\Lambda \in [-1,1]^{n \times m}$ by
\be \label{eq:neww1}
\l_{ij} = \th_{ij} - \psi_{ij} , \fa i , j.
\ee
Thus $\L$ would be the zero matrix if $X$ and $Y$ were independent.
Define
\bd
\nm \L \nm_{1,1} := \max_{j = 1 , \ldots , m} \sum_{i=1}^n | \l_{ij} |
\ed
to be the matrix norm of $\L$ induced by the vector $\ell_1$-norm.
Finally, for a given vector $\bnu \in \R^m$, let $\text{Diag}(\bnu)$ 
denote an $m \times m$ diagonal matrix with the elements of $\bnu$ on
the diagonal.
With these definitions, the following result can be derived.

\begin{theorem}\label{thm:3e}
\cite [Theorem 6] {Ahsen2014} Suppose $X,Y$ are random variables over finite sets $\Abb,\Bbb$ with
joint distribution $\bth$ and marginals $\bmu,\bnu$ respectively.
Then
\be
\phi(X,Y) = \max_j \frac{1}{\nu_j} \sum_i ( \l_{ij} )_+ 
= 0.5 \; \max_j \frac{1}{\nu_j} \sum_i | \l_{ij} | 
= 0.5 \; \nm \Lambda [ {\rm Diag} ( \bnu ) ]^{-1} \nm_{1,1} , \label{eq:3h}
\ee
where $(a)_+ = \max \{ a, 0 \}$ denotes the nonnegative part of a number.
\end{theorem}

Note that Theorem \ref{thm:3e} requires a total of $mn$ algebraic operations
and $m$ comparisons;
therefore $\phi(X|Y)$ can be calculated in polynomial time. 

Another very relevant property of the $\phi$-mixing
coefficient is given next.
\begin{theorem}
\cite [Theorem 9] {Ahsen2014} Whenever $(X \perp Z ) | Y$, the following inequality holds:
\be\label{eq:dpi-phi}
\phi(X|Z) \leq \min \{ \phi(X|Y) , \phi(Y|Z) \} .
\ee
\end{theorem}

Note that (\ref{eq:dpi-phi}) is entirely analogous in appearance to
(\ref{eq:dpi-inf}).
For this reason, we will refer to (\ref{eq:dpi-phi}) as the data processing
inequality (DPI) for the $\phi$-mixing coefficient.

\subsection{The Phixer Algorithm}

With the aid of the above properties, it is possible to design
an algorithm for inferring GINs.
The first step is called ``pruning.''
Start with a ``complete'' network where there is an edge from every
node to every other node.
Then, for each triplet $(X,Y,Z)$ of nodes, check whether Equation
\eqref{eq:dpi-phi} is satisfied.
If so, this means that the edge from node $Z$ to node $X$ can be
deleted, and the resulting network would still be consistent with the
data.
The order in which
the triplets are checked does not affect the final network.
The second step is called ``thresholding,'' and is a consequence of
small sample sizes as are common in cancer data sets.
In reality, we can compute \textit{only an approximation} to
the true number $\phi(X|Y)$ based on a finite number of cellular
measurements.
Consequently, even if the true value were to be zero, the computed
value would not exactly be zero, though it would be small.
Therefore all edges that remain after pruning should have their
weights compared against a threshold, and those whose weight is
smaller than this threshold should be seen as spurious and thus discarded.
We have chosen a value of $0.1$ for the threshold, because
if $\phi(X|Y) < 0.1$,
then ignoring the possible dependence of $X$ on $Y$ leads to a maximum
error of $0.1 = 10$\% in computing various probabilities.
One can play around with this threshold of $0.1$, and the properties
of the resulting networks do not vary very much.
In the actual implementation, we employ bootstrapping, so as to ensure
that the resulting network is very stable;  see Figure \ref{fig:flow}.

Complete details of the implementation of the phixer algorithm are
given in the supplementary material.
A parallel implementation of the Phixer algorithm in C language is available at the web site
\url{https://github.com/nitinksingh/phixer}.

\begin{figure*}[btp]
    \centering
    	\includegraphics[width=0.99\textwidth]{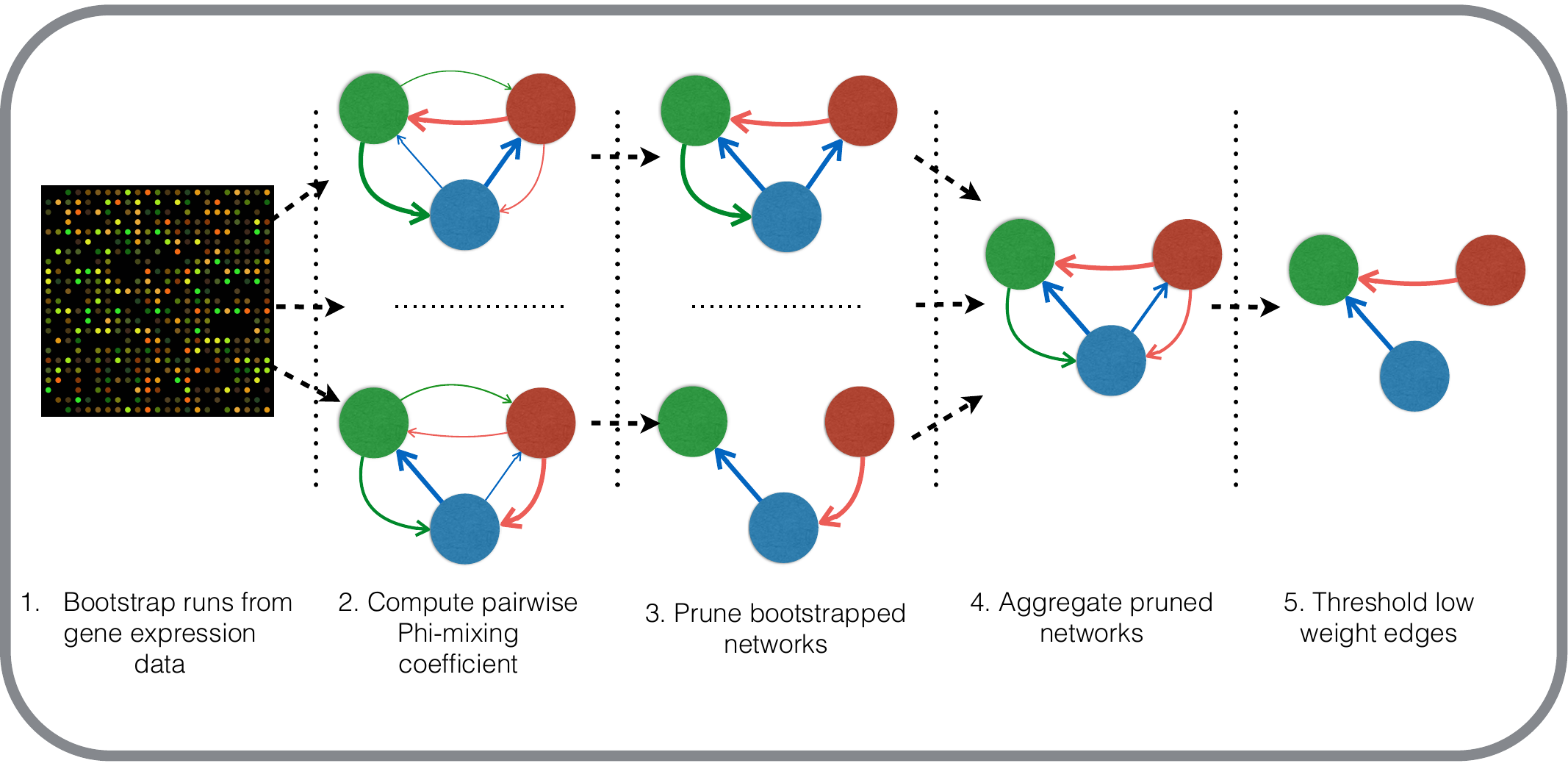}
    \caption{\label{fig:flow}
	An illustration of the phixer algorithm. 
The thickness of an edge illustrates its weight.
The networks are aggregated over several bootstrap runs to avoid over-fitting.
In each run, the phi-mixing coefficient is computed between every pair of genes.
The resulting complete graph is pruned with the data processing inequality.
The aggregated network is then thresholded to retain only high weight edges. 
}
\end{figure*}

\subsection{Properties of the Phixer Generated GINs}

The GINs produced by phixer algorithm have the following features:
\ben
\item The GIN is invariant under any monotone transformation of the data.
In other words, if $f_i : \R \ap \R, i = 1 , \ldots , n$ are any
monotonic functions, then the GIN produced by applying the
algorithm to the original set
$\{ x_{ij} , i = 1 , \ldots , n , j = 1 , \ldots , m \}$ will be exactly
the same as the GIN produced by applying the algorithm to the transformed
data set $\{ f_i (x_{ij}), i = 1 , \ldots , n , j = 1 , \ldots , m \}$.
This feature is useful as the post-processing of raw measurements 
often involves log transformation.
\item The GIN has weighted, directional edges.
\item Each edge in the GIN has a weight between $0$ and $1$.
\item The resulting GIN is permitted to contain cycles, and the edges
are directional.
That is, if $A$ and $B$ are two nodes in the GIN, then it is possible
to have an edge from $A$ to $B$ but not from $B$ to $A$, and it is also
possible to have edges from $A$ to $B$ and from $B$ to $A$, while
the weights are the two edges could be different.
\item The resulting GIN is \textit{strongly connected}; that is, there
is a directed path between every pair of nodes.
\een

\section{Results}

\subsection{Details of the Experiments Performed}

Ideally the performance of a network inference algorithm should be
assessed by comparing the predicted network with the actual network.
Unfortunately, there is no ``ground truth'' for genome-wide networks on a human
scale.
Therefore we analyzed the outcomes of
several experiments on lung cancer, and matched
the predictions from the inferred networks with experimental results.

Specifically, we used the phixer algorithm to construct and characterize
three networks based on the
whole-genome gene-expression data (19,464 genes) of 157 lung cancer cell 
lines and one network from 59 normal lung tissue cell lines.
The lung cancer cohort consisted of
three distinct subtypes, namely: non-small cell lung cancer 
(NSCLC; 106 samples),
small cell lung cancer (SCLC; 40 samples),
and neuroendocrine non-small cell lung cancer
(NE-NSCLC; 11 samples).
Neuro-endocrine NSCLC is viewed as a separate subtype of NSCLC
\cite{Bhattacharjee2001, Jones2004, Augustyn2014}.
About 10\% of the NSCLC cell lines 
show neuroendocrine morphological features 
that are characteristic of the SCLC subtype.
These NE-NSCLC cell lines do in fact show a distinct gene-expression
pattern compared to the rest of the NSCLC cell lines \cite{Augustyn2014}.
However, it is not possible to infer a network based on just 11 samples.
Since NE-NSCLC appears to be closer to SCLC than NSCLC, we grouped 
the 11 NE-NSCLC cell lines with the 40 SCLC cell lines, to
generate a total of 51 samples, which is nevertheless referred to
as the SCLC cohort.
The remainder of the 106 NSCLC cell lines are called the NSCLC cohort.
The normal cell line cohort (59 samples) consists 
of both human bronchial epithelial cell (HBEC) and 
human small airway epithelial cells (HSAEC) cell lines.

These inferred networks were analyzed to see whether they exhibit
scale-free or power law behavior.
Next, these networks were used to predict the essentiality of each
of the 19,000+ genes to the survival of a cancerous cell;
the predictions were compared with the outcomes of
siRNA screening of 19,000+ genes on 11 NSCLC cell lines.
Finally, the immediate ancestors and successors of the
lineage specific oncogenic transcription factor ASCL1
were determined from the inferred networks for SCLC and NSCLC cell lines.
These were compared with the outcome of
a ChIP-Seq experiment to determine putative downstream targets of ASCL1.

\subsection{Phixer Generates Sparse and Scale Free Networks}

\begin{figure*}[btp]
    \centering
    	\includegraphics[scale=0.2]{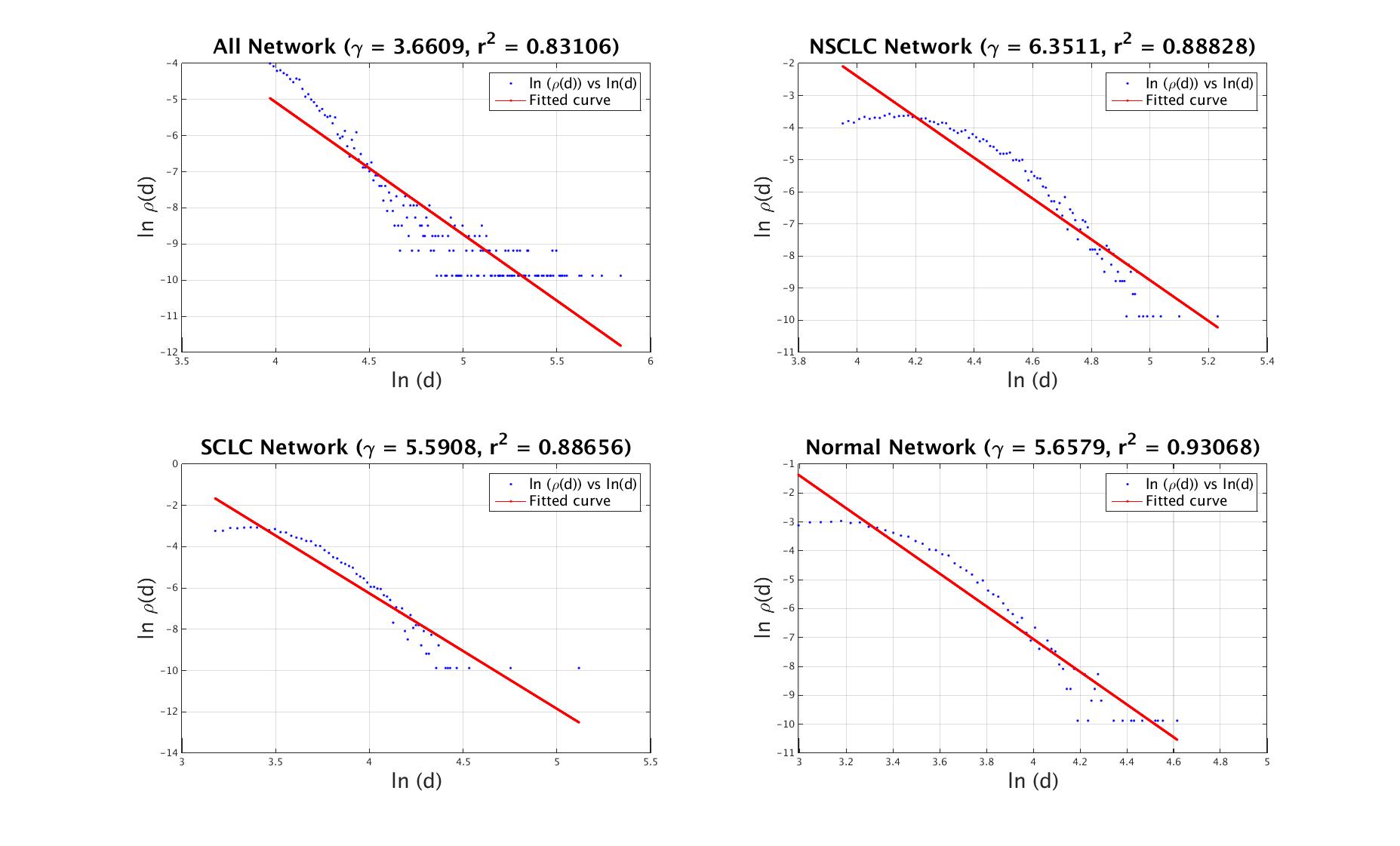}
    \caption{\label{fig:power_law}
Lets $\rho(d)$ denote the fraction of nodes in the GIN that 
have degree $d$.
To test whether the distribution of node degree exhibits
a scale-free or power law distribution,
the quantity $\rho(d)$ is plotted versus
the degree $d$ on a log-log scale,
and  a least square fit to the degree distribution
of all four networks and the corresponding exponents are also shown.
It is evident that in both the networks there are a few very high degree
genes, also referred as ``master regulators'' or ``hubs.''
The fit is after dropping the genes whose total degree falls in the 
bottom $20$ percentile.
All of these networks exhibit scale-free behavior.
}
\end{figure*}

It is widely believed that real biological networks consist of
a few ``master regulators,'' each of which controls hundreds of
other genes, while the non-regulators have relatively few connections
between themselves \cite{Barabasi1999,Newman2002a}.
The phixer-generated networks were analyzed on the basis of two parameters,
namely degree distribution and the clustering coefficient, to determine
how well they conform to this popular notion.
Precise definitions of these two concepts are given in the
Supplementary Material.
But roughly speaking, the degree distribution measures the extent to which
the network under study differs from a ``random'' network.
In addition, we also examined whether the degree distributions in various 
phixer-generated networks
display ``scale-free'' or 
``power law behavior,'' whereby the number of nodes having
a particular degree drops off as a power of the degree itself;
in other words, on a log-log scale the plot looks linear.
As shown in Figure \ref{fig:power_law}, all four of the networks
demonstrate a power law behavior.
The clustering coefficient, as the name suggests, measures
the tendency of nodes to cluster together.
As shown in Figure \ref{fig:power_law}, all four of the networks
demonstrate small clustering coefficients.

\subsection{Gene Connectivity versus its Essentiality}

In this part of the study, we explored whether there is a relationship
between the degree of a gene in the phixer-generated GIN and its
essentiality for cell survival.
It has been suggested that differentially expressed \cite{Wachi2005}
or often mutated \cite{Jonsson2006} genes in cancer have a higher degree
in protein-protein interaction (PPI) networks.
The advantage of PPI networks is that the edges
correspond to regulatory relationships that have been experimentally
validated.
However, PPI networks have an element of a ``self-fulfilling prophecy''
in the sense that ``interesting'' genes are studied far more than
``uninteresting'' genes, with the result that ``interesting genes''
have more proven interactions.
In short, one finds what one is looking for, a classic instance
of finder's bias.
In contrast, the phixer algorithm ``treats all genes equally'' while
generating the GIN.
Therefore, if a gene degree in the phixer-generated GIN is substantially
higher than that of another gene, this is a consequence of the data
and not finder's bias.

Specifically, we investigated whether there is a relationship
between the degree of a gene in the phixer-generated network,
and its essentiality for cell survival.
The natural hypothesis is that higher degree genes are more 
essential to the cell than lower degree genes.
To test this hypothesis, a subset of 11 NSCLC cell lines from
the 106 samples used to infer the NSCLC network
were subjected to genome-wide siRNA screening with two different
siRNA libraries: Ambion and Dharmacon, resulting 
a set of 22 siRNA screening measurements.
There were 18,534 common genes between the genome wide siRNA screening 
and the gene-expression measurements.
The siRNA screening assay quantifies the intra-cellular ATP 
concentration for as a measure of the consequence of each gene depletion event on 
cell survival.
The raw ATP concentration values were
converted to $z$-scores by subtracting the mean and dividing by
the standard deviation.
The resulting $z$-scores are also referred to as the lethality scores.
A higher (i.e., more positive or less negative)
$z$-score for a gene implies that the gene has relatively
low effect on cell survival.
On the other hand, a lower (more negative) $z$-score  implies that the
gene is essential for cell survival.
Therefore, as per the hypothesis, one would expect a negative correlation
between
the lethality score and the total degree of a gene.
The significance of the resulting correlation is quantified against 
the null hypothesis that
there is no negative correlation between the degree and lethality score.
Therefore, the $p$-values of the significance are calculated against the 
one-sided permutation distributions.

The degree of a gene in the GIN is indicative of its centrality in
the network only at a coarse level, and not at a very fine grain level.
To illustrate, a gene whose node degree is 400 is certainly more central
than one whose node degree is 100; but it cannot be said to be more
central than one whose node degree is 390.
Therefore some amount of \textit{grouping} genes by their degrees is warranted.
To study the relation at a coarse level,
we sorted the genes by their total degree and then 
aggregated them into groups consisting of $k$ genes each,
where $k$ is varied from $1, 10, 20, \ldots, 1000$.
Note that the case $k=1$ corresponds to computing the correlation between lethality of an 
individual gene and its total degree, without any grouping of genes.
Now the correlation coefficient
between the average leathality score and the average degree 
of these groups was computed.
This kind of aggregate analysis was performed to study the robustness of 
the correlation over a wide range of group sizes.
Figure \ref{fig:essential} summarizes these results 
showing that 14 out of 22 cell-line screens have a statically significant
($P$-value $< 0.05$) correlation.
Evidently, for these 14 cell-lines, the negative correlation is robust to 
variations in the group size, with only a small number of outliers.

\begin{figure*}[tbp]
	\centering
      \includegraphics[scale=.45]{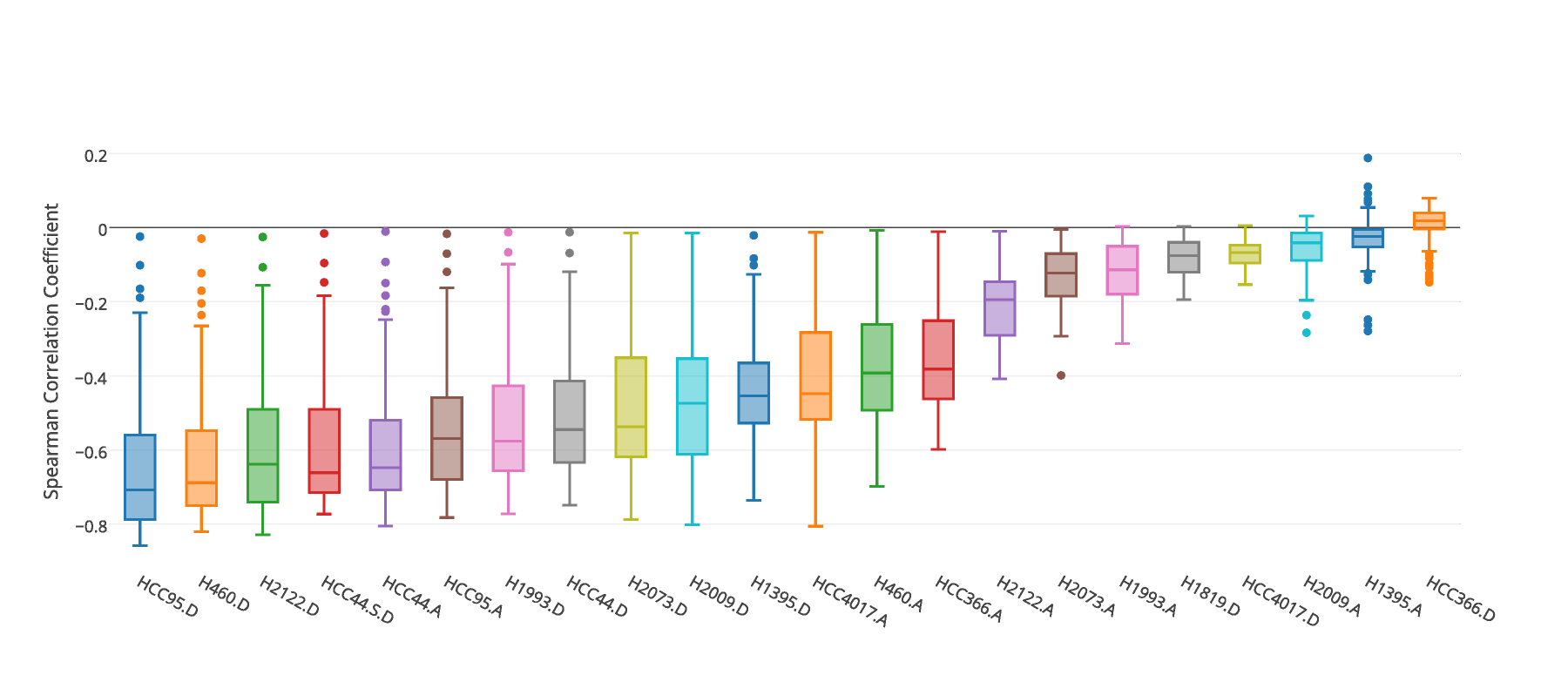}
\caption{
The distribution of the Spearman correlation coefficient between the total degree of a gene versus its siRNA lethality score for 22 siRNA screening experiments. The cell line name suffix'.A' and '.D' respectively represent the screening libraries Ambion and Dharmacon. To check the robustness of the correlation, genes were grouped by their total degree and the correlation between the average degree of the group and the average leathality score was computed. The group size was varied from 1 to 1000 in a step size of 10 genes i.e., $1, 10, 20, \ldots, 990, 1000.$}
\end{figure*}
\begin{figure*}
\centering
      \includegraphics[scale=.45]{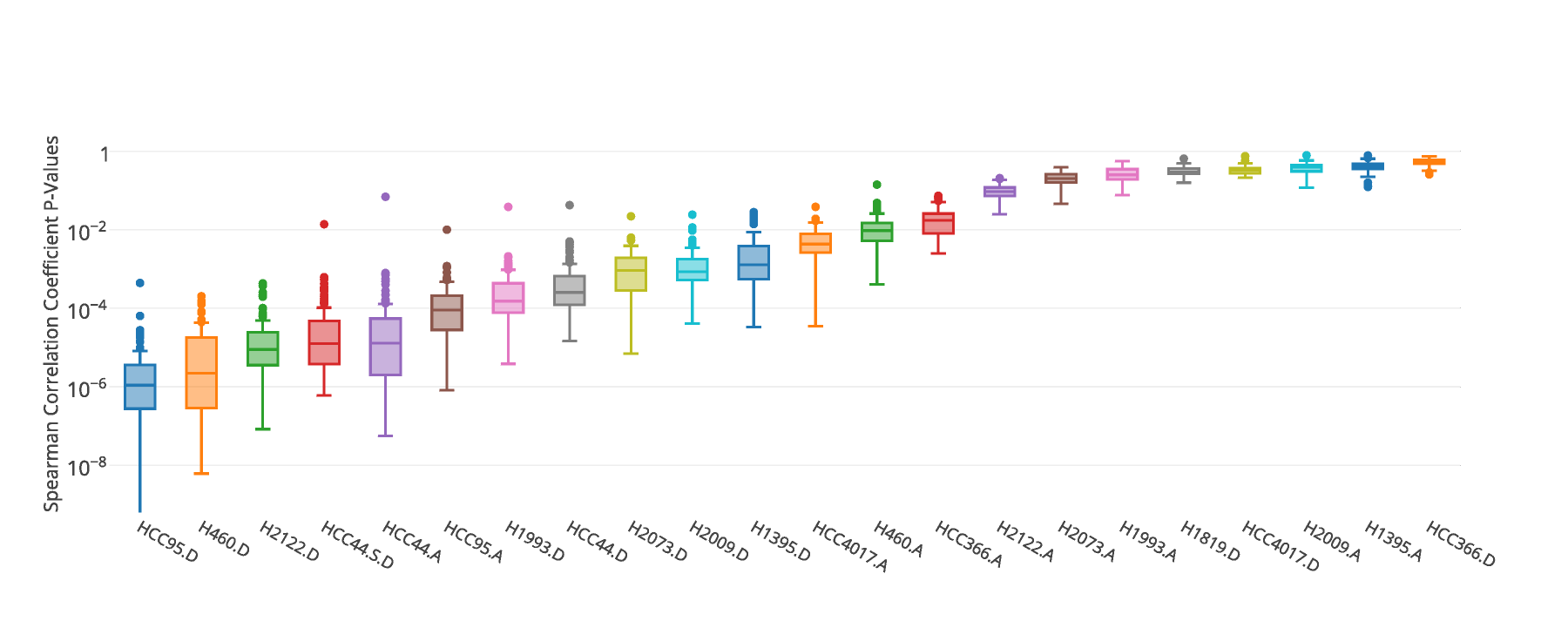}
    
\caption{The distribution of the P-Values of the Spearman correlation coefficients between the total degree of a gene versus its siRNA lethality score. The P-Values (single-tailed) were computed against the null hypothesis that there is non-negative correlation between the average degree and the average lethality score.}
    
    \label{fig:essential}
\end{figure*}

\subsection{Phixer Infers Context-Specific Networks Accurately}

\subsubsection{Neighborhood of ASCL1 in the Inferred GIN is Enriched
for ChIP Target Genes}

The predictive ability of the phixer algorithm was assessed by comparing the 
immediate neighborhood of the transcription factor achaete-scute homolog 1
(ASCL1) in the phixer-produced GIN with its putative targets
obtained via an orthogonal ChIP-Seq experiment.
The lineage specific transcription factor ASCL1 is required for the
differentiation of pulmonary cells \cite{Borges1997}, and the initiation
and survival of SCLC cells are dependent on its expression \cite{Jiang2009}.
Further, it has been recently reported that ASCL1 is a lineage specific oncogene
that is also required for the survival of NE-NSCLC cells \cite{Augustyn2014}.
A ChIP-Seq experiment was performed on two ASCL1(+) SCLC cell 
lines (NCI-H889 and NCI-H2107), while another two ASCL1(-) cell lines
(NCI-H524 and NCI-H526) were used as controls.
In the ChIP-Seq experiment, high quality peaks were analyzed for 
their functional association with the \textit{cis}-regulatory regions with
GREAT \cite{McLean2010} yielding a set of potential ASCL1 targets.
This gene set was further refined by filtering out genes whose
expression in ASCL1(+) cell lines was less than 1.5 times that in
the control.
This resulted in a set of 226 putative ASCL1 target genes
that were used
for the validation of the phixer generated networks.
In the inferred networks, any gene that has a direct edge from the ASCL1 is 
treated as its direct target. 
Note that the putative targets generated from ChIP-Seq studies represent
direct transcriptional binding, while edges in the GIN represent 
``informational'' interactions that may or may not correspond to
``regulatory'' binding.

The set of genes in the GIN that were downstream of ASCL1 was
analyzed for enrichment of the ChIP-Seq generated putative target genes.
The null hypothesis was that the downstream genes in the GIN were simply
generated at random; therefore the $P$-value was computed using
the hypergeometric distribution.
In the SCLC network based on 51 samples, out of 37 downstream neighbors
of ASCL1 in the phixer-generated GIN, 6 were in the list of 226 ChIP-Seq
genes, out of 19,464 genes in the network.
The associated $P$-value of this happening by chance is $4\times 10^{-6}$.
In the network based on all 157 lung cancer samples, out of 119
downstream neighbors of ASCL in the phxer-generated GIN, 31 were from
the ChIP-Seq list, for a $P$-value of $1.8\times 10^{-33}$.
In contrast, in the network based only on 106 NSCLC samples,
out of the 27 downstream neighbors of ASCL1 in the phixer-generated GIN,
only 1 was from the ChIP-Seq target list, for a $P$-value of $0.27$.
This is consistent with the known fact that ASCL1 has a significant
role \textit{only} in SCLC (and its near-cousin, neuro-endocrine
NSCLC), but \textit{not} in NSCLC.
Thus it would have been surprising had the neighborhood of ASCL1 been
enriched with ChIP-Seq genes also in the NSCLC network.
Figure \ref{fig:ascl1_enrich} (a) illustrates the neighborhood of ASCL1 in the all lung cancer
network.

\subsubsection{ASCL1 Enrichment Results are Robust}

As briefly mentioned in the introduction, phixer-generated networks are 
aggregated over many bootstrap runs and then finally undergo a 
thresholding step during which low weight edges are deleted. 
When two genes do not interact in reality,
the theoretical phi-mixing coefficient should be zero.  
However, when the coefficient is estimated on the basis of a finite number of
samples, the computed coefficient may not exactly equal zero, but will be small.
Therefore, any edge whose computed mixing coefficient is smaller 
than $\approx 0.1$ can be deleted on the basis that the nonzero weight is 
an artifact of the small sample size.
In the various networks, it is observed that a threshold higher than
$\approx 0.1$ causes the network to become disconnected.
The threshold value ($\approx 0.1$) is a user-defined parameter,
and its actual 
value is determined from the distribution of the edge weights 
of the network; see SM Figure 1.  
The second parameter of the phixer algorithm is the number of bootstrap runs. 
 
To study the effect of parameters on the ASCL1 enrichments, 
the number of bootstrap runs was varied from 100 to 300 in a 
step size of 50, while the threshold value was varied from 0 to $\approx 0.1$,
and the enrichment $P$-values are computed. 
Figure 3(b) shows that the ASCL1 targets are enriched in all lung cancer
network ($P$-value  $< 10^{-30}$)
and SCLC network ($P$-value $< 10^{-5}$) network, but not in the NSCLC network 
($P$-value $> 0.1$) for the range of parameter values studied. 
Applying a threshold higher than $\approx 0.1$ may cause these networks to 
become disconnected, and is hence not recommended. 
Figure \ref{fig:ascl1_enrich} (b) also shows that the ASCL1 enrichment 
results are qualitatively the same for varying numbers of bootstrap runs. 
Since the run time of the algorithm scales linearly with the number of
bootstrap 
runs, it is suggested to run the phixer algorithm only for 100 bootstrap runs. 
Together these results establish that the ASCL1 enrichment results 
are robust to the variation in phixer parameter values.

\begin{figure*}
    \centering
\includegraphics[width=0.99\textwidth]{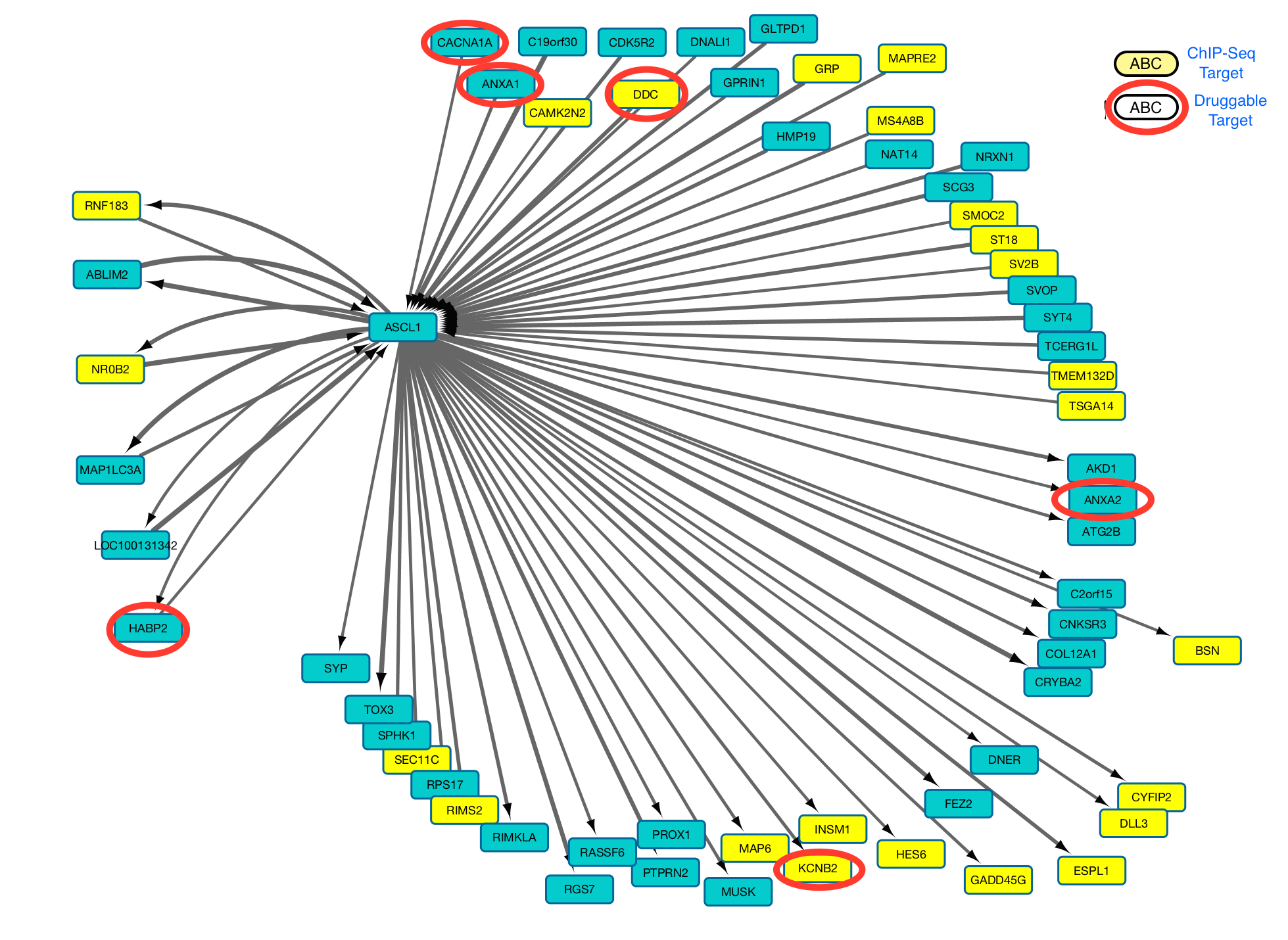}
\caption{
The upstream and downstream neighbors of ASCL1 in all lung cancer network. }
\end{figure*}

\begin{figure*}
\centering
\includegraphics[width=0.99\textwidth]{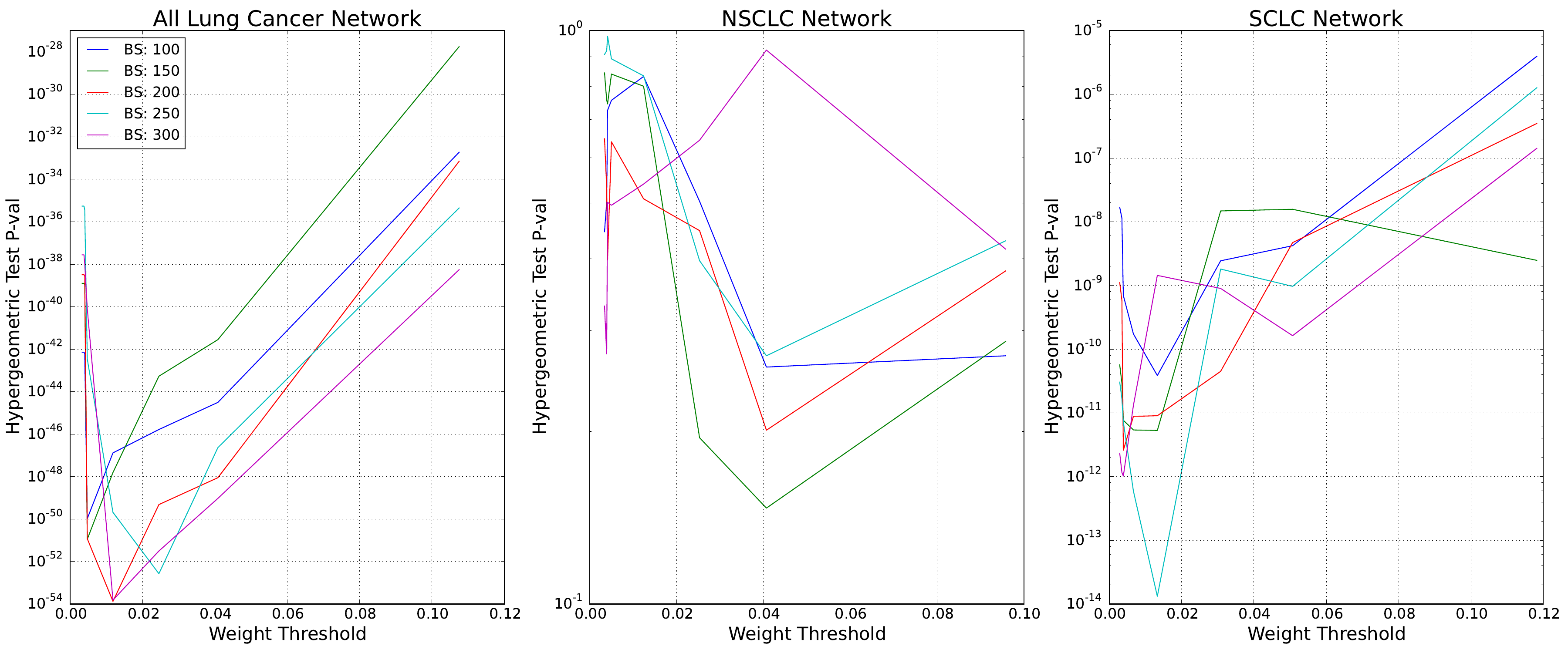}
\caption{Robustness of ASCL1 enrichment $P$-values with variation in parameters namely, the number of bootstrap runs and the threshold value.}
\label{fig:ascl1_enrich}
\end{figure*}

\subsubsection{Phixer Compares Favorably with ARACNE}

Algorithm for the Reconstruction of Accurate Cellular Networks (ARACNE)
\cite{Margolin2006} is a widely used GIN inference algorithm that 
builds upon previous methods \cite{Butte2000, Faith2007} based on computing
the mutual information between genes.
The precursor to ARACNE is the influence network \cite{Butte2000},
which is an undirected graph consisting of edges
between nodes $i$ and $j$ whenever the mutual information between genes $i$
and $j$ exceeds a user-defined threshold.
Influence networks are extremely dense because they do not distinguish
between direct and indirect influence.
In ARACNE, one begins with the influence network, and then
``prunes'' it using
the data processing inequality \cite{Cover2006}.
In this manner, ARACNE prunes an edge unless its presence is necessary
to explain the data at hand.
This algorithm has been validated in the context of human B-cells \cite{Basso2005}.
The main shortcoming of ARACNE, and other methods such as CLR that
use mutual information as a criterion, is that the resulting network
is always \textit{undirected}, because the mutual information between
two random variables is a symmetric quantity.
Nevertheless, at present ARACNE is perhaps the most widely used
algorithm for constructing mammalian GINs at a genome-wide scale.

The open source Linux version of the ARACNE software was downloaded
from the official webpage\footnote{\url{http://wiki.c2b2.columbia.edu/califanolab/index.php/Software/ARACNE}} and used for all the computations below.
We generated all three lung cancer networks with the ARACNE algorithm with
the default 
options with the exception of enabling the DPI
($e = 0$, default is 1 i.e.,\ no DPI) 
and running for 100 bootstraps ($r = 100$, default is 1 to save time).
The enrichment of the ASCL1 neighborhood in the ARACNE-generated network
for the 226 ChIP-Seq putative targets was computed using the
hypergeometric distribution, as earlier.
In the SCLC network generated using 51 samples, 2 out of 16 neighbors
were ChIP-Seq, for a $P$-value of $0.14$.
In the all lung cancer network based on 157 samples, 6 out of 13 neighbors
were ChIP-seq, for a $P$-value of $3.6\times 10^{-9}$.
In the NSCLC network based on 106 NSCLC samples, none of the
40 neighbors was a ChIP-Seq target, for a $P$-value of 1.
Note that since the ARACNE networks are undirected the ASCL1 targets consist
both upstream and downstream targets.
Recall that the corresponding $P$-values for the phixer networks were 
$4\times 10^{-6}$ (6 out of 37), $1.8\times 10^{-33}$ (31 out of 119) 
and $0.27$ (1 out of 27), if only genes that are downstream of
ASCL1 in the GIN are used to compute the $P$-value.
To make a like-to-like comparison with ARACNE, which does not provide
any directional information of the interactions, we recomputed
the $P$-values using both upstream as well as downstream genes.
The resulting
$P$-values were $1.6\times 10^{-5}$ (6 out of 47), $3.2\times 10^{-51}$ (46 out of 164) and $0.46$ (1 out of 54) 
for the SCLC, all lung cancer and NSCLC networks respectively.
This shows that phixer provides a more accurate 
and context specific neighborhood for ASCL1 compared to
the ARACNE algorithm.
We observe that the ASCL1 neighborhood predicted by the ARACNE network 
contains far fewer genes than one would expect for an oncogene
that is critical to the initiation and survival of neuroendocrine 
lineage lung cancers.
In contrast, the number of downstream genes of ASCL1 in the phixer-generated
GIN, using the suggested threshold of 0.1 for the weight, is quite
reasonable for an oncogene.
Moreover, the phixer algorithm provides a threshold parameter
to control the sparsity of the resulting network.

\section{Discussion}

As gene interaction networks (GINs) can shed light on fundamental
biological processes and
disease mechanisms, systematic construction of GINs from high-throughput
gene expression data
is one of the important problems in systems biology. 
In this paper, we have presented a new algorithm, named phixer, to infer 
GINs from gene expression data.
In contrast with existing algorithms,
the phixer algorithm generates a network whose edges are
\textit{directed and weighted}, and is also permitted to contain cycles.
This is in contrast to information-based methods which lead to undirected
graphs, and Bayesian methods which lead to acyclic graphs.
The proposed method was utilized to construct and contrast three lung cancer
networks and one network from normal tissue samples.
An analysis of these networks revealed that they display 
a power law behavior (also known as ``scale-free'' behavior)
and clustering properties;
this matches widely-held expectations that real world biological networks
have these properties.
It is widely believed that ``hub'' genes, that is, genes that interact
with many other genes, are more essential to cell survival than genes
with fewer interactions.
This hypothesis was tested on 11 lung cancer cell lines using two different
reagents to perform siRNA knockdown experiments.
The lethality score of each gene on each cell line was correlated with
its node degree in the inferred GIN.
In 14 out of 22 experiments, there was a statistically significant correlation.
Finally, the putative target genes of the transcription factor ASCL1,
as determined by a ChIP-seq experiment, were tested against the predicted
upstream and downstream genes of ASCL1 in the inferred GIN.
The neighborhood was enriched with the putative target genes with an
extremely low $P$-value.
A comparison of phixer with widely used GIN construction algorithm
ARACNE shows that the ASCL1 neighborhood is enriched with ChIP-Seq targets
to a far higher extent in the network constructed using
the phixer algorithm, compared to the network constructed using ARACNE.
In addition, the run time of phixer is about half that
of ARANCE (see Methods section).

The ability to infer
context specific GINs may prove to be a valuable tool to investigate 
biological processes that have a clinical potential.
First, the observation that there is a significant correlation between
the degree of a gene in the reverse-engineered GIN and its lethality score
can be of clinical value, by identifying genes that can be suppressed to
reduce cancer cell proliferation, whether or not the genes have
a known role in the onset or progression of the disease.
This relationship could potentially be exploited to seek
genes whose activation (or inactivation)
is lethal to tumor cells
but relatively innocuous to the normal cells.
Second, identifying druggable upstream regulators of an oncogene
is valuable when the oncogene might not itself be directly druggable.
Of course, any such hypotheses based on the inferred GIN must then
be validated experimentally.

The performance of
any network GIN inference method, including phixer, is governed by
the number of available samples. 
Because the method requires the estimation of the dependence between
two random variables, ideally the number of samples should in excess of 100.
Also, it must be recognized that the edges in the network represent 
``informational'' relationships
that may not necessarily correspond to physical interaction or causation.
Nevertheless, we believe that the context specific GINs 
inferred using the phixer algorithm will be useful to shed light on
interesting biological hypotheses. 


\section*{Acknowledgements}

The authors thank Prof.\ John Minna, Prof. Jane Johnson, Dr.\ Alex Augustyn and Dr. Mark Borromeo 
for the gene-expression and ChIP-Seq data.

NS, MEA, SM and MV were supported by the National Science Foundation
under ECCS Awards \#1001643 and \#1306630;
the Cancer Prevention and Research Institute of Texas (CPRIT) under
grants RP110595 and RP140517; the Jonsson Family Graduate Fellowships; and
the Cecil~H.\ and Ida Green Endowment at the University of Texas at Dallas.
HSK was supported by the
National R\&D Program for Cancer Control, Ministry of Health \& Welfare,
Republic of Korea \#1420100, and in part by the CPRIT training grant RP101496.
MAW was supported by the Welch Foundation award I-1414,
and NIH grants CA197717, and CA176284.


\bibliographystyle{IEEEtran}

\bibliography{library,biology-refs}

\end{document}